\documentclass[aps,prl,twocolumn,superscriptaddress,showpacs,floatfix]{revtex4}

\usepackage{graphicx}
\usepackage{dcolumn}
\usepackage{bm}

\newcommand{\pbar}{$\overline{p} $}
\newcommand{\pbarhelium}{\pbar He${}^+$}
\newcommand{\pbarheliumion}{\pbar He${}^{++}$}

\newcommand{\pbhe}{\pbarhelium\ }
\newcommand{\nuHF}{$\nu_{\text{HF}}$}

\newcommand{\nuHFp}{$\nu_{\text{HF}}^+$}
\newcommand{\nuHFm}{$\nu_{\text{HF}}^-$}
\newcommand{\nuSHF}{$\nu_{\text{SHF}}$}
\newcommand{\nuSHFp}{$\nu_{\text{SHF}}^+$}
\newcommand{\nuSHFm}{$\nu_{\text{SHF}}^-$}
\newcommand{\nuMW}{$\nu_{\text{MW}}$}
\newcommand{\us}{$\mu$s}

\newcommand{\Se}{$\vec{S}_e$}
\newcommand{\Lp}{$\vec{L}_{\overline{p}}$}
\newcommand{\Sp}{$\vec{S}_{\overline{p}}$}
\newcommand{\mup}{$\vec{\mu}_{\overline{p}}$}
\newcommand{\mue}{$\vec{\mu}_e$}

\newcommand{\Dthexp}{$\Delta_{\text{th-exp}}$}
\newcommand{\dexp}{$\delta_{\text{exp}}$}
\newcommand{\Fp}{$F^+$}
\newcommand{\Fm}{$F^-$}
\newcommand{\fp}{$f_+$}
\newcommand{\fm}{$f_-$}

\begin{document}

%
\title{Hyperfine structure of antiprotonic helium revealed by
    a laser-microwave-laser resonance method}

\author{E.~Widmann}
\author{R.S.~Hayano }
\author{T.~Ishikawa}
\author{J.~Sakaguchi}
\author{T.~Tasaki}
\author{H.~Yamaguchi }
\affiliation{Department of Physics, University of Tokyo, 7-3-1 Hongo,
Bunkyo-ku, Tokyo 113-0033, Japan}
\author{J.~Eades}
\author{M.~Hori}
\affiliation{CERN, CH-1211 Geneva 23, Switzerland}
\author{H.A.~Torii}
\affiliation{Institute of Physics, University of Tokyo, Komaba,
Meguro-ku, Tokyo 153-8902, Japan}
\author{B.~Juh{\'a}sz}
\affiliation{Institute of Nuclear Research of the Hungarian Academy of
Sciences, H-4001 Debrecen, Hungary}
\author{D.~Horv{\'a}th}
\affiliation{KFKI Research Institute for Particle and Nuclear Physics,
H-1525 Budapest, Hungary}
\author{T.~Yamazaki}
\affiliation{RI Beam Science Laboratory, RIKEN, Wako, Saitama 351-0198,
Japan}

\date{\today}

\begin{abstract}
Using a newly developed laser-microwave-laser resonance method, we
observed a pair of microwave transitions between hyperfine levels of
the $(n,L)=(37,35)$ state of antiprotonic helium. This experiment
confirms the quadruplet hyperfine structure due to the interaction of
the antiproton orbital angular momentum, the electron spin and the
antiproton spin as predicted by Bakalov and Korobov. The measured
frequencies of \nuHFp $=12.89596 \pm 0.00034$ GHz and \nuHFm $=12.92467
\pm 0.00029$ GHz agree with recent theoretical calculations on a level
of $ 6 \times10^{-5}$.
\end{abstract}


\maketitle

In this paper we report the first observation of microwave-induced
transitions between magnetic substates of antiprotonic helium and use
the results to determine its quadruplet hyperfine splitting to better
than 1 MHz (relative precision $\sim 3\times 10^{-5}$). Antiprotonic
helium is an exotic three-body system consisting of a helium nucleus,
an antiproton, and an electron (\pbar$-e^--$He$^{2+} \equiv $
\pbarhelium). It has a series of highly excited metastable states
(lifetime $\sim$ \us) with principal quantum number $n$ and angular
momentum quantum number $L$ in the range 33--39, which have been
extensively studied by laser spectroscopy (see
\cite{Iwasaki:91,Yamazaki:93} and a comprehensive review
\cite{Yamazaki:01}). In the most recent experiments performed at the
CERN Antiproton Decelerator (AD), the wavelengths of several
laser-induced transitions of the antiproton in \pbhe were measured with
a relative accuracy of $1.3\times10^{-7}$, leading to a CPT test
limiting any relative difference in the masses and charges of the
proton and antiproton to $6\times10^{-8}$ \cite{Hori:01}. A step
further in the precision spectroscopy of antiprotonic helium is the
investigation of its magnetic {\em hyperfine structure}, i.e., the
level splitting caused by the magnetic interaction of the  \pbar\
orbital angular momentum \Lp, the electron spin \Se, and the \pbar\
spin \Sp. To the leading order, the electron in \pbhe is in the ground
state with a spin magnetic moment $\vec{\mu}_e = g_e \, \mu_{\text{B}}
\, \vec{S}_e$. The \pbar\ magnetic moment, on the other hand, consists
of an orbital part and a spin part $\vec{\mu}_{\overline{p}} =
(g^{\overline{p}}_{\ell} \, \vec{L}_{\overline{p}} +
g^{\overline{p}}_{\text{s}} \, \vec{S}_{\overline{p}}) \,
\mu_{\overline{\text{N}}}$. The orbital $g$-factor,
$g^{\overline{p}}_{\ell}$, defines the relation between the \pbar\
orbital magnetic moment and the anti-nuclear magneton
$\mu_{\overline{\text{N}}} =Q_{\overline{p}} \hbar / (2
M_{\overline{p}})$. Its value is usually implicitly taken to be one,
but this assumption has never been tested experimentally before, either
for the proton or the antiproton.

Due to the large angular momentum of \pbarhelium, the dominant
splitting arises from the magnetic interaction of \Se\ with \Lp. Thus,
the coupling of \mup\ with \mue\ creates a doublet (called here {\em
hyperfine (HF)} splitting) with  $\vec{F} = \vec{L}_{\overline{p}} +
\vec{S}_e$ ($F^- = L - 1/2$ and $F^+ = L + 1/2$). The interaction of
the \pbar\ spin magnetic moment with the other magnetic moments splits
each sublevel $F^+$ and $F^-$ into a still finer doublet, and if
$\vec{J} = \vec{F} + \vec{S}_{\overline{p}}$ is the total angular
momentum, these further sublevels are associated with its components
$J^{-+} = F^- + 1/2 = L, J^{--} = F^- - 1/2 = L  -1, J^{++} = F^+ +1/2
= L+1,$ and $ J^{+-} = F^+ - 1/2  = L$. We refer to this as {\em
superhyperfine (SHF)} splitting.

\begin{figure}[t]
\includegraphics[width=8.6cm]{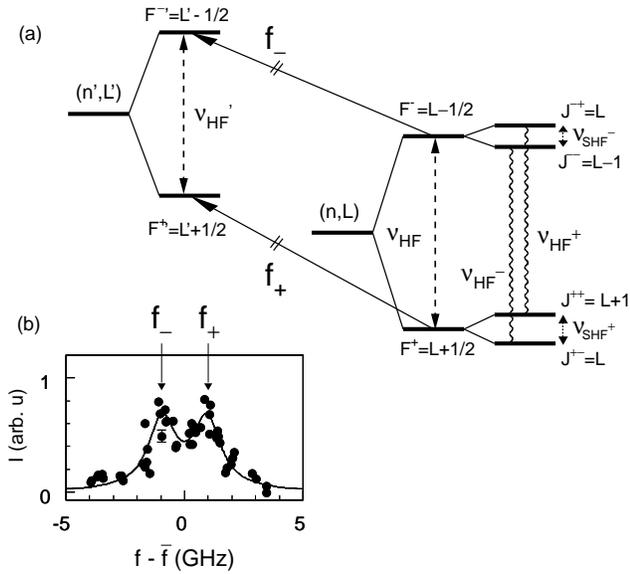}%
\caption{\label{fig:leveldiag} (a) Schematic view of the splitting of a
\pbhe state and observable laser transitions from the $F^{\pm}$ levels
of a $(n,L)$ state to a daughter state $(n',L')$ (arrows). Wavy lines
denote allowed magnetic transitions associated with an electron spin
flip. (b) Laser scan of the 726.1 nm transition with $(n,L)=(37,35)$
and $(n',L')=(38,34)$ performed at the AD ($\overline{f} =
($\fp$+$\fm$)/2)$.}
\end{figure}

The theoretical framework of the hyperfine structure of \pbhe has been
established by Bakalov and Korobov \cite{Bakalov:98} who showed that
the SHF splitting resulting from the one-body spin-orbit coupling of
the antiproton (\Lp$\cdot$\Sp ), and the scalar (\Sp$\cdot$\Se) and
tensor
((\Sp$\cdot$\Se)$-3$(\Sp$\times$\Lp)$\cdot$(\Se$\times$\Lp)/$[L(L+1)]$)
spin-spin couplings,  yield the level order as shown in
Fig.~\ref{fig:leveldiag}(a) due to an approximate cancellation of the
latter two spin-spin coupling terms. Typical values for the HF
splitting of metastable states are \nuHF\ = 10 -- 15 GHz, while the SHF
splitting is almost 2 orders of magnitude smaller (\nuSHF$^\pm$ = 150
-- 300 MHz). In a previous experiment at LEAR we indeed observed a
doublet splitting of the unfavoured laser transition
$(n,L)=(37,35)\rightarrow(38,34)$ at $\lambda=726.1$ nm, the two
sub-lines being separated by $\Delta = f_+ - f_- =$ \nuHF (initial)$-$
\nuHF(final) $=1.75\pm0.05$ GHz \cite{Widmann:97}, in agreement with
the theoretical value \cite{Bakalov:98}.   Although the SHF splitting
is too small to be resolved with the pulsed laser system used in our
experiments, it causes a small splitting of the hyperfine transition
into two components, $\nu_{\text{HF}}^+$ and $\nu_{\text{HF}}^-$ for
$J^{-+} \leftrightarrow J^{++}$ and $J^{--} \leftrightarrow J^{+-}$,
respectively, as shown in Fig. 1(a). In order to determine the two
hyperfine frequencies of the $(37,35)$ state on the MHz level directly,
we developed the laser-microwave-laser resonance method described
below.

\begin{figure}[t]
  \includegraphics[width=7.5cm]{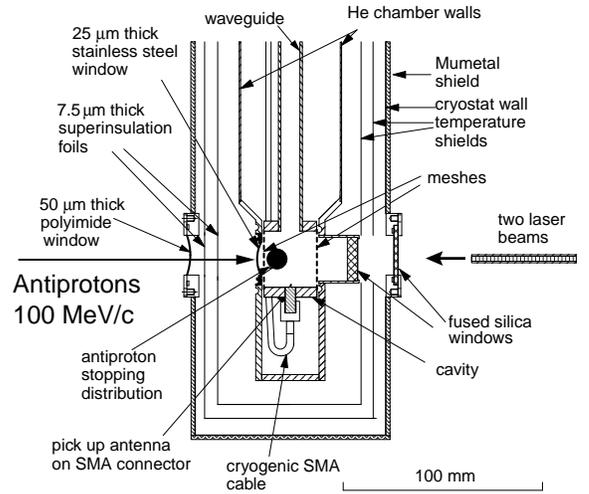}%
  \caption{\label{fig:setup}Side view of the cryostat holding the
  helium target chamber and microwave cavity. The antiprotons enter the
  helium chamber from left and stop inside a cylindrical microwave
  cavity. Two overlapping laser beams come from opposite direction along
  the cavity cylinder axis, and the microwave radiation is supplied
  through a rectangular wave guide from top. Two {\v C}erenkov counters
  not shown in the drawing were placed on both sides of the cryostat
  where there is no window.}
\end{figure}

The experiment was performed at the Antiproton Decelerator (AD) of
CERN, which delivered pulses of 2 -- 4 $\times 10^7$ \pbar\ of 200 ns
length (FWHM) with a momentum of 100 MeV/$c$ (5.3 MeV kinetic energy).
One such pulse was extracted from the AD every $\sim 2$ minutes and
stopped in helium gas (see Fig.~\ref{fig:setup}) at a temperature of
6.1 K and pressures of 250 or 530 mbar (number densities $3.0$ or $6.7
\times 10^{20}$ cm$^{-3}$, respectively).  As described in further
detail in Ref.\ \cite{Hori:01}, the time spectrum of delayed
annihilations (ADATS, for Analog Delayed Annihilation Time Spectrum)
was recorded in a digital oscilloscope as the envelope of the output of
photomultipliers (PMTs) connected to  two {\v C}erenkov counters
through which the antiproton annihilation products passed. Because 97\%
of antiprotons stopped in helium annihilate promptly (within
picoseconds), the PMTs were turned off by a gate pulse until $\sim 440$
ns after the center of the AD pulse.

\begin{figure}[t]
  \includegraphics[width=8.6cm]{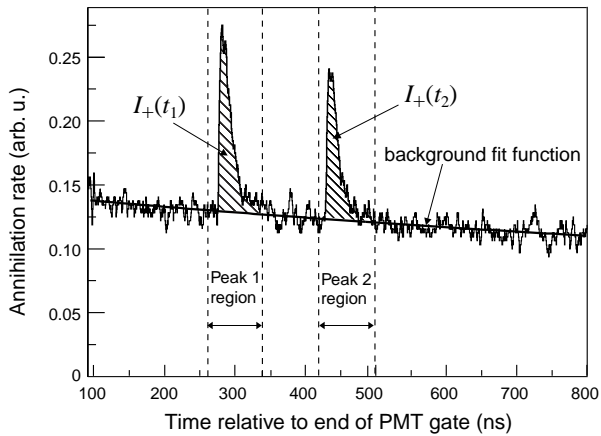}%
  \caption{\label{fig:ADATS} Part of ADATS with two successively
  applied laser pulses of frequency \fp. The background fit
  function and the two peak regions that are excluded from the
  background fit are also shown. $I_+(t_1)$ and $I_+(t_2)$ stand for
  the hatched areas under the two spikes.}
\end{figure}

The wavy lines in Fig.~\ref{fig:leveldiag}(a) represent allowed M1
transitions (flipping \Se\ but not \Sp ) which can be induced by
microwave radiation. All the HF levels are initially nearly equally
populated. In order to create a population asymmetry which is needed to
detect a microwave transition, a laser pulse stimulating a transition
from a metastable ($\tau \sim$ \us) state to a short-lived ($\tau
\lesssim $ 10 ns) state can be used. When the \pbar\ is excited to the
short-lived state, the \pbhe undergoes an Auger transition to a
\pbarheliumion\ ion which is immediately destroyed via collisional
Stark-effect in the dense helium medium followed by annihilation of the
\pbar\ with a nucleon. An on-resonance laser pulse therefore superposes
a sharp spike onto the ADATS (cf. Fig.~\ref{fig:ADATS}) whose area is
proportional to the population of the metastable state at the time of
the arrival of the laser pulse.

The laser-microwave-laser resonance method utilizes the following
sequence: {\em i)} a laser pulse tuned to one of the doublet lines
(e.g.\ \fp\ in Fig.~\ref{fig:leveldiag}(a)) preferentially depopulates
the \Fp\ over the \Fm\ doublet. {\em ii)} The microwave pulse is
applied; if it is  resonant with either \nuHFp\ or \nuHFm, it transfers
population from the \Fm\ to the \Fp\ doublet.  {\em iii)} A second
laser pulse at frequency \fp\  measures the new population of \Fp\
after the microwave pulse.

The laser light of $\lambda =$ 726.1 nm was produced by a commercial
dye laser pumped by a Nd:YAG laser which was triggered synchronously
with the arrival of the antiproton pulse. The line width of the dye
laser with intra-cavity etalon was 0.6--0.8 GHz. A resonance scan (cf.
Fig.~\ref{fig:leveldiag}(b)) showed a doublet structure with a
separation of $\Delta = f_+ - f_- = 1.8 \pm 0.1$ GHz, in agreement with
our earlier observation at LEAR \cite{Widmann:97}. The two sequential
laser pulses were obtained by dividing the output of the dye laser and
delaying one part by multiple reflections. In this way a maximum delay
of 160 ns could be obtained without seriously degrading the laser beam
spot quality. The laser power density was adjusted to achieve an
optimum compromise between the depletion efficiency, i.e., the
efficiency of depopulation of the \Fp\ doublet, and the power
broadening which reduces the  population asymmetry.

The Rabi frequency for the allowed microwave transitions is given by
$\Gamma_{av} = (1/(4 \sqrt{2})) g_e \mu_B B_1$, where $B_1$ is the
amplitude of the oscillating magnetic field. From this formula as well
as from detailed numerical simulations
\cite{Sakaguchi:MSnotit,Sakaguchi:01} it follows that in order to
induce an M1 transition  in the 160 ns time difference between the two
laser pulses, a $B_1$ of several gauss is needed. To apply the
microwave radiation, we constructed a cylindrical cavity for \nuMW\
$\sim 12.9$ GHz (diameter 28.8 mm, length 24.6 mm) resonating in the
TM$_{110}$ mode.  The ends of the cylinder were covered by metal meshes
with a transparency of 85\% permitting the \pbar\ and the two laser
beams to enter from opposite directions. The cavity was immersed in the
low-temperature helium gas, and the microwave radiation was applied
through a rectangular wave guide. An external triple stub tuner (TST)
was used to tune the central frequency and $Q$-value of the cavity,
thus allowing the microwave frequency to be scanned over a range of
$\sim 200$ MHz corresponding to $\sim 1.5$\% of the resonance
frequency, while keeping $Q$-values of $\sim 2700$ \cite{Sakaguchi:01}.
The resonance characteristics of the cavity were measured with a vector
network analyzer (VNA) both in reflection and in transmission mode. The
microwave pulse to induce the electron spin-flip transition was
generated by amplifying the VNA output by a pulsed travelling wave tube
amplifier (TWTA).

A measurement cycle consisted of the following computer-controlled
steps. Before arrival of the antiprotons, the three stubs of the TST
were set to obtain the desired resonance frequency and $Q$-value, and
the cavity characteristics were verified using the VNA. The VNA was
then tuned to this frequency, and upon arrival of the \pbar\ beam, we
fired the two laser pulses and the microwave pulse into the target. The
ADATS of two {\v C}erenkov counters showing two laser spikes each (cf.
Fig.~\ref{fig:ADATS}) were recorded.

\begin{figure}
  \includegraphics[width=8.6cm]{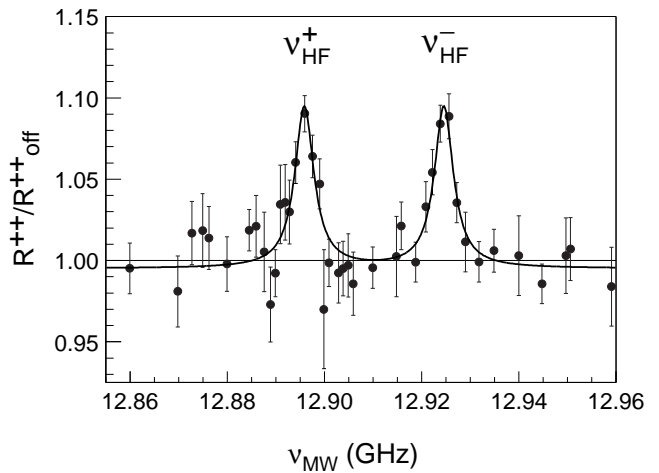}%
  \caption{\label{fig:MWscans} Average of all microwave scans showing
  clearly two resonance lines as predicted. The width of the lines of
  $\sim 5$ MHz corresponds to $4\times10^{-4}$ of the central
  frequency.}
\end{figure}

We collected data during five 8-hour periods. At the beginning of each
period, the laser alignment and depletion efficiency were verified. We
then collected approximately 180--290 AD shots, stepping through 10--30
microwave frequencies in cycles. The microwave power was set to the
optimum value (15 W) corresponding to an oscillating field strength of
$B_1=7$ gauss inside the cavity  as obtained from computer simulations.
Each cycle included 1--3 points at very low microwave power for use as
``microwave-off'' reference points.

For each ADATS of one data taking period, we fitted a background
function (a sum of two exponentials) to the time spectra in the range
from 60 ns after the PMT was turned on to 500 ns after the first laser
peak, excluding the two peak regions from the fit as shown in
Fig.~\ref{fig:ADATS}.  We extracted the difference between the observed
spectrum and the fit function in the two peak windows as shown in
Fig.~\ref{fig:ADATS}. The peak areas $I_+(t_i)$ are proportional to the
population of the \Fp\ doublet at time $t_i$. In order to reduce
systematic effects such as fluctuations in the overlap of laser and
\pbar\ beams or the \pbar\ intensity, which might affect both peaks
identically, we calculated the ratio $R^{++} \equiv I_+(t_2)/I_+(t_1)$
and plotted it against \nuMW. The individual data sets each showed two
peaks at the theoretically predicted positions for \nuHFp\ and \nuHFm.
The off-resonance value of $R^{++}$ agreed with the points taken at
extremely low microwave power, but both varied from data set to data
set. This results from uncontrollable systematic effects associated
mainly with laser misalignments.

Each data set was fitted by a sum of two Lorentzian functions with
identical width and amplitude, plus a constant background. The results
for \nuHFp\ and \nuHFm\ as well as the width of the Lorentzians agreed
within the error bars and did not show any dependence on the target
density. This is consistent with theoretical arguments presented by
Korenman \cite{Korenman:priv} that the shift of the line centres with
density is very small, and that the collisional broadening at our
densities is of the order of MHz.

We therefore normalized the individual data sets to the microwave-off
value $R^{++}_{\text{off}}$, and averaged points within 0.7 MHz  to
give the final spectrum shown in Fig.~\ref{fig:MWscans}. The chosen
averaging interval was smaller than the line width $\gamma_{\text{obs}}
= 1/(2\pi \; 160 \; \text{ns}) = 1.0$ MHz corresponding to the 160-ns
observation time window.  The natural width of the $(37,35)$ state of
$\gamma_{(37,35)} = 0.12$ MHz is smaller than $\gamma_{\text{obs}}$,
while the measured line width $\gamma_{\text{exp}} = 5.3 \pm 0.7$ MHz
was significantly larger. This may be due to collisions, to the
inhomogeneity of the magnetic field over the stopping distribution of
\pbar, or to the fact that the many substates with magnetic quantum
numbers $m = -J \ldots J$ each have different Rabi frequencies.

\begin{table}[b]
\caption{\label{tab:HFS} Experimental values for the HF transition
frequencies of the state $(37,35)$ in GHz compared to theoretical
results. The relative experimental error \dexp\ and the difference
\Dthexp $\equiv (\nu_{\text{th}}-\nu_{\text{exp}})/\nu_{\text{exp}}$
are given in ppm.}
\begin{center}
\begin{ruledtabular}
\begin{tabular}{llD{.}{.}{-1}lD{.}{.}{-1}}
   &  \multicolumn{1}{c}{\nuHFp}& \multicolumn{1}{c}{\dexp\ }
   & \multicolumn{1}{c}{\nuHFm}& \multicolumn{1}{c}{\dexp\ }  \\
   &  \multicolumn{1}{c}{(GHz)} & \multicolumn{1}{c}{(ppm)}
   & \multicolumn{1}{c}{(GHz)}& \multicolumn{1}{c}{(ppm)}\\
  \hline
   Exp. &$12.895\,96(34)$ &27& $12.924\,67(29)$& 23\\ \hline
   &  & \multicolumn{1}{c}{\Dthexp\ }  &
   & \multicolumn{1}{c}{\Dthexp\ } \\ \hline
  BK \protect\cite{Bakalov:98} &$12.895\,97$  &0.6 &12.923\,94
  &$-57$\\
  KB \protect\cite{Korobov:01} &$12.896\,3462$ &30 &$12.924\,2428$
  &$-33$\\
  YK \protect\cite{Yamanaka:00}&$12.898\,977$ &234 &$12.926\,884$
  &$171$\\
  K\protect\cite{Kino:01notit} &$12.896\,07391$
  & 8.6 &$12.923\,96379$ &$-55$
\end{tabular}
\end{ruledtabular}
\end{center}
\end{table}

The final results for \nuHFp\ and \nuHFm\ obtained from fitting two
Lorentzians plus a constant background to the spectrum of
Fig.~\ref{fig:MWscans} are presented in Table~\ref{tab:HFS} and
compared to recent theoretical calculations by two groups. The
theoretical values for for \nuHFp\ and \nuHFm\ distribute over a much
wider range ($\sim$ 100 ppm) than the laser transition energies ($\sim$
0.1 ppm) calculated by the same groups (see comparison in Ref.\
\cite{Hori:01}), reflecting a higher sensitivity of the hyperfine
coupling terms to the details of the wave functions involved.
Nevertheless, the experimental values are in excellent agreement with
the results of Korobov and Bakalov (both their initial values BK
\cite{Bakalov:98} and their most recent ones KB \cite{Korobov:01})  as
well as the latest values of Kino {\it et al.} (K \cite{Kino:01notit}).

In summary, we have established a laser-microwave-laser resonance
method and succeeded in observing  two microwave transitions, \nuHFp\
and \nuHFm, in antiprotonic helium. The experiment has fully confirmed
the presence of a quadruplet structure originating from the hyperfine
coupling of \Lp, \Se, and \Sp, as predicted by Bakalov and Korobov
\cite{Bakalov:98}. The agreement of the experimental values \nuHFp\ and
\nuHFm\ with the most updated theoretical values KB and K is about $6
\times 10^{-5}$ or better, on the level of the accuracy of the
calculations. These do not include contributions of order
$\alpha^2\approx 5\times10^{-5}$ or higher. Presently, the experimental
error is about $3\times10^{-5}$, slightly exceeding the theoretical
precision. The microwave resonance frequencies, \nuHFp\ and \nuHFm, are
primarily related to the dominant \pbar\ orbital magnetic moment, which
is probed by the aid of the large electron magnetic moment. As implied
in the definition of \mup\ (paragraph 1), \nuHFp\ and \nuHFm\ depend on
the spin part $g^{\overline{p}}_s$ as well as on
$g^{\overline{p}}_\ell$. This dependence is however too weak to permit
a precise determination of $g^{\overline{p}}_s$, which would require a
direct measurement of \nuSHFp\ or \nuSHFm. Thus, the excellent
agreement between experiment and theory can be interpreted as an
experimental measurement of the antiprotonic orbital $g$-factor with a
relative precision of $\sim 6 \times 10^{-5}$. We note that no
experimental value exists for $g^p_\ell$ for the proton because no
atoms with an orbiting proton exist in the world of ordinary matter.

We thank Dr. Fritz Caspers (PS Division, CERN) for invaluable help in
the design of the microwave cavity and circuits as well as D. Bakalov,
V.I. Korobov, Y. Kino and N. Yamanaka for many helpful discussions and
for providing us with their results prior to publication. The support
by the CERN cryogenic laboratory as well as the help of Mr.\ K.\ Suzuki
and Dr.\ H.\ Gilg is acknowledged. The work was supported by the
Grant-in-Aid for Creative Basic Research (10NP0101) of Monbukagakusho
of Japan, the Hungarian Scientific Research Fund (OTKA T033079 and
TeT-Jap-4/02), and the Japan Society for the Promotion of Science.

\newcommand{\SortNoop}[1]{} \newcommand{\OneLetter}[1]{#1}
  \newcommand{\SwapArgs}[2]{#2#1}



\end{document}